# Mid- and long-wavelength infrared computational ghost spectroscopy

Linzhen He[1], Han Wu[1]*, Wei Deng[1], Mingzhu She[2], Bo Hu[1], Zunwang Bo[3], Shensheng Han[3], Weili Zhang[2]*, Houkun Liang[1]* and Goëry Genty[4]*

China | [1]School of Electronics and Information Engineering, Sichuan University, Chengdu, Sichuan 610064,

E-mail: hanwu@scu.edu.cn and hkliang@scu.edu.cn

China |[2]Key Lab of Optical Fiber Sensing & Communications, University of Electronic Science & Technology of China, Chengdu, Sichuan 611731,

E-mail: wl_zhang@uestc.edu.cn

China |[3]Key Laboratory for Quantum Optics, Shanghai Institute of Optics and Fine Mechanics Chinese Academy of Science, Shanghai 201800,

[4]Finland |Laboratory of Photonics, Tampere University, FI-33014 Tampere,

E-mail: goery.genty@tuni.fi

## Abstract

Spectral-domain ghost imaging enables high-resolution spectroscopy with a single-pixel detector. The technique does not rely on spectrally resolved detectors, which makes it inherently robust against turbulence and particularly adapted to weak-light conditions. These features are very attractive for spectral imaging in the mid-infrared region which hosts numerous molecular absorption features but lacks highly sensitive detectors. The implementation of spectral ghost imaging in the mid-infrared has however been limited by the absence of suitable light sources and detectors capable of generating and measuring spectral fluctuations in real time. Here, we demonstrate spectral-domain computational ghost imaging in the mid-infrared based on a nonlinear frequency downconversion scheme. Pre-programmed spectral patterns imposed on broadband light at 1.5 μm using a programmable spectral filter are transferred into the mid-infrared through difference-frequency generation in a nonlinear crystal. This enables computational ghost spectroscopy with a spectral resolution of 0.62 $cm^{-1}$ using a single-pixel mid-infrared detector. The method is flexible, broadly applicable and, as proof of concept, we demonstrate ghost spectroscopy in mid-wavelength infrared (3.25–3.47 μm) and long-wavelength infrared (6.9–7.45 μm) bands using the nonlinear frequency conversion in chirped-poling lithium niobate and $ZnGeP_2$ crystals, respectively. Our approach provides a new avenue for mid-infrared spectroscopy, remote sensing and spectral imaging.

## 1 | Introduction

Ghost imaging (GI) is an indirect imaging technique originally demonstrated in the spatial domain[1–3], allowing to reconstruct an image by correlating the spatial intensity distribution of a reference beam that does not interact with the object and the spatially integrated intensity of a test beam illuminating the object. Computational ghost imaging (CGI) where the physical reference beam is replaced with preprogrammed patterns generated by a digital micromirror array or a spatial light modulator[4,5] has emerged as a powerful alternative approach, allowing for high-resolution image reconstruction using only a single-pixel detector. In addition, CGI provides inherent robustness against noise and turbulence, as well as enhanced security because it relies solely on integrated intensity detection[6,7].

More recently, exploiting space-time duality, GI has been extended to the temporal and spectral domains[8,9], leading to the concept of temporal ghost imaging and spectral ghost imaging (SGI), techniques which have enabled new approaches for ultrafast waveform detection[10–15] and spectroscopic measurements[16–19], respectively. Similar to spatial GI, SGI uses a broadband incoherent light source as a spectral probe and high-resolution spectral measurements can be realized by correlating real-time, spectrally resolved fluctuations of the source with the integrated intensity transmitted through a spectral object[17]. Computational SGI has also been demonstrated using preprogrammed spectral filters applied to broadband light or spectral-temporal modulation based on dispersive Fourier transform[20–23]. Comparative experiments with direct detection have shown that computational SGI preserves the robustness of ghost imaging against turbulence and enables accurate absorption spectroscopic measurements in low-light and scattering conditions[20].

Up to now, demonstrations of SGI have been essentially limited to the near-infrared (NIR) region below 2 μm, where high-sensitivity detectors and low-loss fiber components are readily available. There is a significant interest in extending the technique into the mid-infrared (MIR) region which hosts the fundamental vibrational absorption bands of most organic and inorganic molecules. Indeed, compared to NIR spectroscopy, MIR absorption spectroscopy exhibits much stronger absorption strengths and provides richer spectral signatures with higher molecular specificity. These features make MIR spectroscopy a key experimental characterization technique across many fields, including gas detection, environmental monitoring, and remote sensing[24–28]. MIR absorption spectroscopy is commonly performed using Fourier-transform infrared spectroscopy[29], tunable-source frequency scanning[30,31], dual-comb spectroscopy[32–34], and up-conversion spectrometry[35,36]. However, the lack of highly sensitive MIR detectors still limits the performance of direct spectral measurements. In addition, open-path configurations are generally prone to turbulence, scattering, and spectral distortion, which can degrade measurement accuracy significantly. Extending SGI to the MIR could overcome these limitations by reducing the detector requirements and enhancing robustness to environmental perturbations, but the lack of MIR-compatible sources and detectors for generating and measuring spectral fluctuations has limited progress in this direction.

Here, we introduce and experimentally demonstrate computational spectral ghost imaging in the mid-infrared using a nonlinear frequency down-conversion approach. In our scheme, preprogrammed spectral patterns are imprinted onto a broadband 1.5-μm source with a programmable spectral filter and subsequently transferred to the MIR via difference-frequency generation with a single-frequency near-infrared laser in a nonlinear crystal. We achieve broadband (3.25–3.47 μm) and high-resolution (0.62 $cm^{-1}$) ghost spectroscopy using a single-pixel MIR detector and a chirped-poling lithium niobate (CPLN) crystal. The versatility of the method is further demonstrated by extending it to the 6.9–7.45 μm range with a $ZnGeP_2$ (ZGP) crystal. This frequency down-conversion computational SGI concept paves the way for high-resolution spectroscopy at wavelengths where sensitive array detectors remain unavailable, extending the reach of ghost imaging into the MIR and THz domains.

## 2 | Principle of computational SGI

In computational spectral ghost imaging, a single-pixel MIR detector records the integrated intensity of light that has been spectrally modulated by preprogrammed patterns and transmitted through a target object with particular spectral transmission (or reflection). The transmission spectrum of the object is then reconstructed by correlating the known spectral patterns with the measured integrated intensities over $N$ distinct realizations. In order to reduce the reconstruction time, we employ a Walsh–Hadamard set for the preprogrammed spectral patterns. Assuming the spectral transmission of the target object is represented by $S = (S_1, S_2, \ldots, S_k)$, where $S_k$ denotes the transmission in different frequency bins, and the preprogrammed spectral patterns are denoted $\boldsymbol{H}_i$ ($i = 1, 2, \ldots, N$) where $N$ is the total number of realizations, the spectral intensity measured by an integrating MIR detector can be expressed as

$$\boldsymbol{B} = \boldsymbol{HS}. \tag{1}$$

In principle, $\boldsymbol{H}$ is an $N \times N$ Walsh-ordered Hadamard matrix composed of elements -1 and 1, but because the detected spectral intensity is non-negative, the negative elements are removed by constructing two complementary Hadamard matrices: $\boldsymbol{H}_{io}$ where all –1 values are replaced by 0 and $\boldsymbol{H}_{ie}$ representing the inverse pattern. The Walsh-ordered Hadamard matrix is then derived from those as

$$H_i = H_{io} - H_{ie}, \tag{2}$$

and the corresponding integral spectral intensities can be rewritten by

$$B_i = B_{io} - B_{ie}. \tag{3}$$

The reconstructed spectral ghost image corresponding to the spectral transmission profile of the target object is then retrieved from the second-order correlation between each Hadamard pattern $\boldsymbol{H}_i$ and the measured integrated intensity $B_i$:

$$S_i' = \langle (B_i - \langle B \rangle)((I_i - \langle I \rangle) \rangle, \tag{4}$$

where $\langle \cdot \rangle$ denotes averaging over $N$ realizations.

## 3 | Results

### 3.1 | Experimental setup of MIR computational ghost spectroscopy

The experimental setup is illustrated in **Figure 1**. A broadband amplified spontaneous emission (ASE) source centered at 1.5 μm is spectrally modulated over the 1542–1561.2 nm range (19.2 nm spectral bandwidth) using a programmable spectral filter (WaveShaper 1000A) loaded with Walsh–Hadamard patterns. The modulated light is then amplified to ~200 mW by an erbium-doped fiber amplifier (EDFA) and directed into a frequency down-conversion module, where it is combined with a single-frequency ytterbium-doped fiber laser operating at 1064 nm as shown in **Figure 2**a (3 W optical power) using a wavelength division multiplexer (WDM). Both beams are focused into a 5-cm-long chirped-poling lithium niobate (CPLN, HC Photonics) crystal specifically designed with a linearly chirped period from 30 to 30.35 μm to provide broadband phase matching for difference-frequency generation (DFG). This configuration accurately transfers the preprogrammed spectral patterns from the near-infrared to the mid-infrared, generating an idler centered at 3.4 μm with a −3 dB bandwidth of about 92 nm and an average power of ~80 μW (see Figure 2b and Figure 2c). A long-pass filter with anti-reflection coating blocks residual NIR light. The spectrally encoded MIR light is then transmitted through the target object (a MIR band-pass filter or a 0.05 mm polystyrene film) and detected by a single-pixel MIR photodetector (Thorlabs PDAVJ5, 2.7–5.0 μm, 1 MHz bandwidth). The transmission spectrum of the object is reconstructed by correlating the measured single-pixel intensities with the preprogrammed spectral patterns applied to the NIR filter.

We first verified the fidelity of the spectral transfer from 1.5 to 3.4 μm by modulating the ASE source with a 32-dimensional Walsh-ordered Hadamard matrix (see Figure 2d, and also Note1 in the Supplementary Material) and measuring the corresponding idler spectra at 3.4 μm using an optical spectrum analyzer (AQ6377, 0.1 nm resolution). The recorded spectra displayed in Figure 2e show a one-to-one correspondence between the applied NIR patterns and the MIR outputs, confirming high-fidelity mapping of the spectral information through DFG.

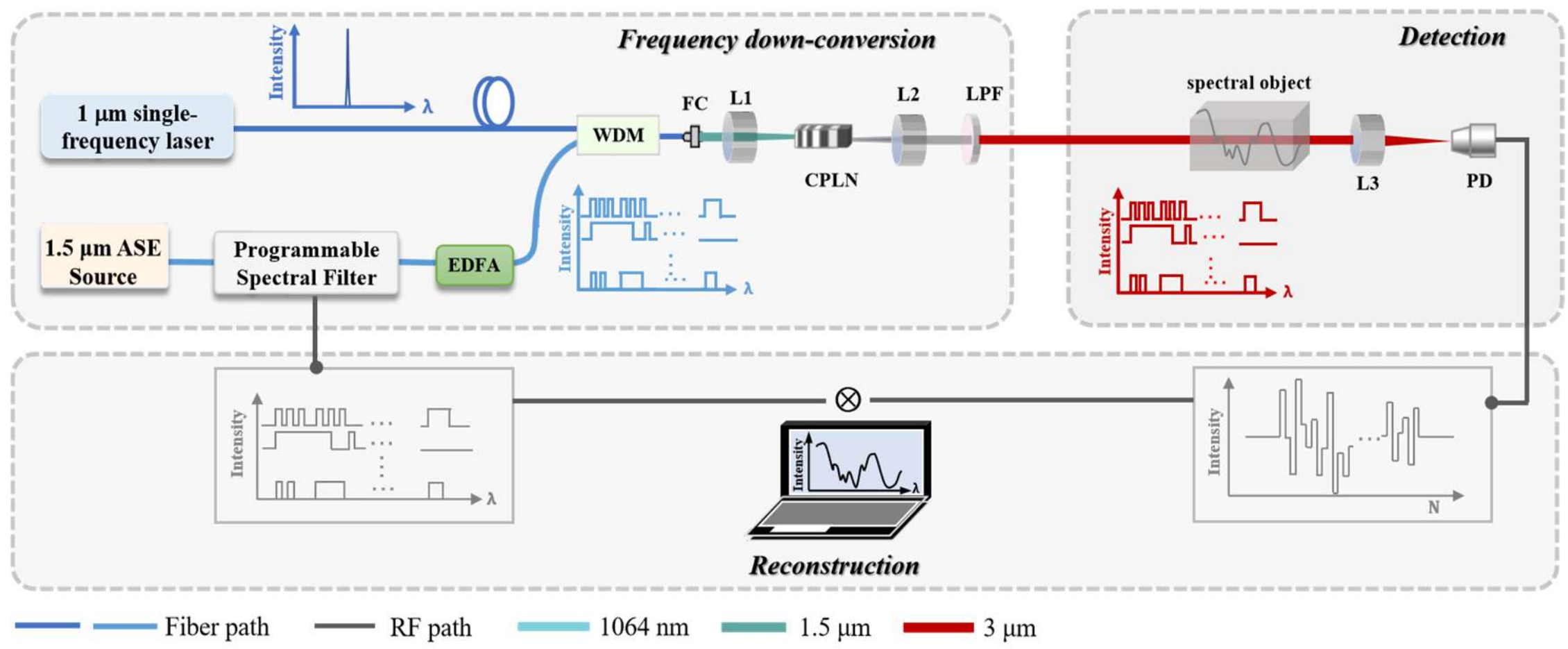

**Figure 1.** Schematic of the mid-infrared ghost spectroscopy system based on frequency down-conversion computational spectral ghost imaging. Walsh–Hadamard patterns are loaded onto a programmable spectral filter to modulate broadband 1.5 μm amplified spontaneous emission (ASE) light over the 1542–1561.2 nm range. After amplification to 200 mW by an erbium-doped fiber amplifier (EDFA), the encoded 1.5 μm patterns are transferred to a 3.4 μm idler via difference-frequency generation with a 1064 nm single-frequency laser in a chirped-poling lithium niobate (CPLN) crystal. The transmitted 3.4 μm light passing through the spectral object is detected by a single-pixel mid-infrared detector, and the spectral information is reconstructed by correlating the applied Walsh–Hadamard patterns with the measured integrated intensities. FC: fiber collimator; L: lens; WDM: wavelength division multiplexer; LPF: long-pass filter.

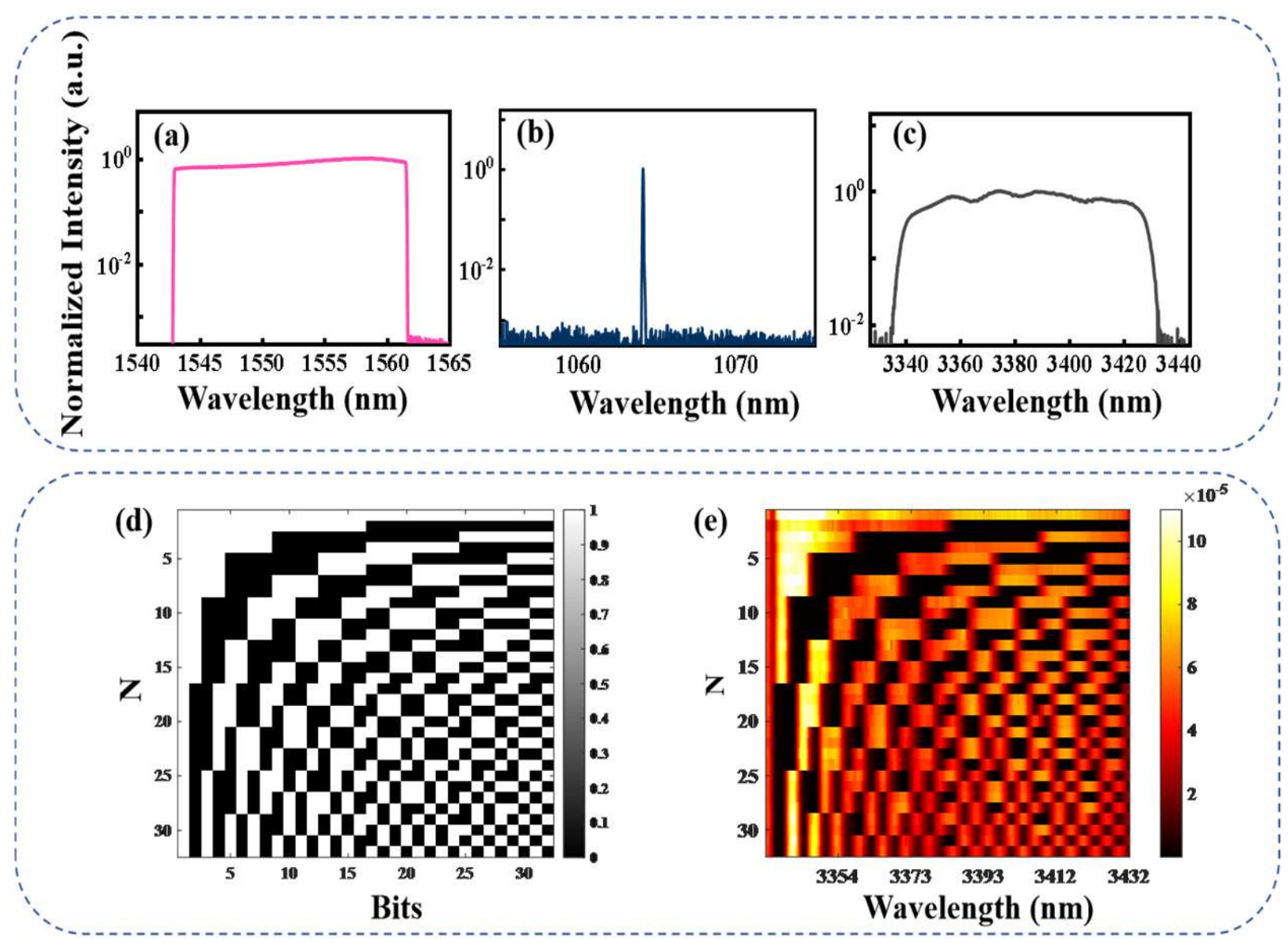

**Figure 2.** Frequency down-conversion of preprogrammed spectral patterns. (a) Spectrum of the 1064 nm single-frequency pump laser. (b) Spectrum of the broadband 1.5 μm signal. (c) Spectrum of the mid-infrared idler generated by difference-frequency generation. (d) 32 Walsh–Hadamard matrix used to spectrally modulate the 1.5 μm broadband light. (e) Corresponding 32 idler spectra at 3.4 μm measured with a high-resolution optical spectrum analyzer.

### 3.2 | MIR ghost spectroscopy based on frequency down-conversion

We next validate the approach by retrieving the transmission spectrum of a 20 nm bandwidth bandpass filter centered at 3380 nm using the computational ghost spectroscopy scheme. The blue dotted lines in **Figures 3**a–**3**d show the reconstructed BPF spectra obtained when the 1.5 μm light was modulated over the spectral range of 1542-1561.2 nm with 16-, 32-, 64-, and 128-dimensional Walsh-ordered Hadamard patterns, respectively. For reference, the shaded area in Figure 3 shows the bandpass filter transmission spectrum measured directly with an optical spectrum analyzer (OSA) with 0.1 nm resolution. One can see that the MIR ghost spectroscopy result obtained with a 16-dimensional Hadamard matrix [Figure 3a] already reproduces well the directly measured spectrum with a spectral resolution of 5.18 cm$^{-1}$. Increasing the matrix dimension and thereby the number of probing spectral patterns progressively enhances the resolution to 2.59, 1.29, and 0.62 cm$^{-1}$ [Figures 3b–3d], while preserving excellent agreement with the reference spectrum. However, a gradual degradation of the signal-to-noise ratio (SNR) is observed as the matrix dimension increases. This behavior reflects the characteristic trade-off between spectral resolution and SNR found in ghost imaging schemes[10] where higher-dimensional matrices provide finer resolution but at the expense of lower SNR and longer acquisition times. The matrix dimension can therefore be flexibly chosen to balance resolution, SNR, and measurement speed depending on the application.

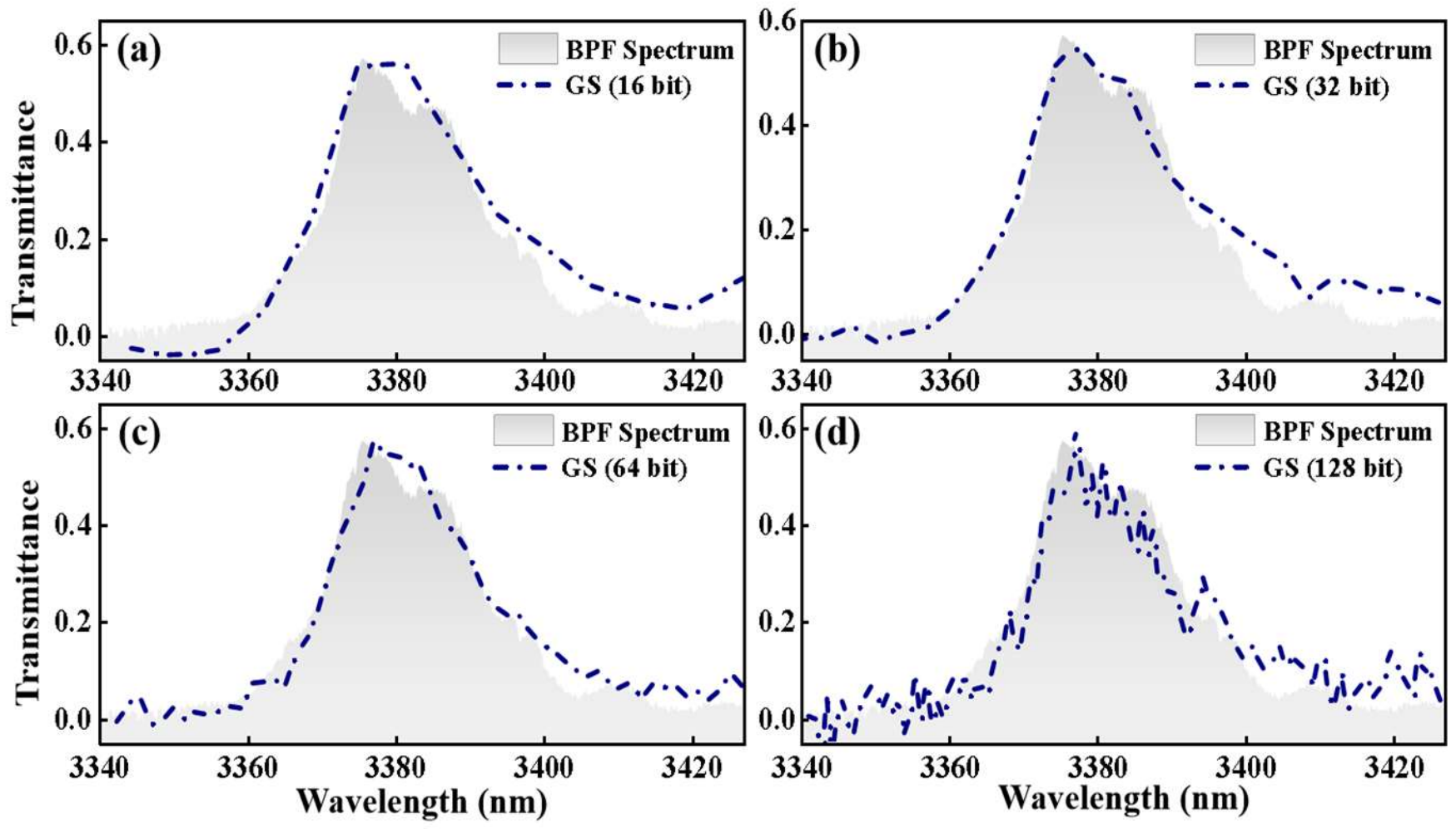

**Figure 3.** Experimental reconstruction of the transmission spectrum of a bandpass filter using mid-infrared computational ghost spectroscopy. (a–d) Reconstructed transmission spectra (blue dotted lines) obtained when the 1.5 µm light was modulated with 16-, 32-, 64-, and 128-dimensional Walsh-ordered Hadamard patterns over the 1542–1561.2 nm range, respectively. The directly measured transmission spectrum of the bandpass filter, recorded with an optical spectrum analyzer, is shown as a shaded area.

To further quantify the trade-off between the number of measurement realizations and reconstruction fidelity, we analyzed the retrieved spectra as a function of the number of probing spectral patterns. Specifically, using a 64-dimensional Hadamard matrix the 1.5 µm light was sequentially modulated with 16, 24, 32, and 64 different patterns and the reconstruction accuracy was evaluated using the peak signal-to-noise ratio (PSNR) between the retrieved spectrum G from the SGI and the directly measured reference spectrum D defined as

$$\mathrm{PSNR} = 10\log_{10}\left[\frac{\mathrm{MAX_D}^2}{\frac{1}{k}\sum_1^k (G_i - D_i)^2}\right], \tag{5}$$

Where K is the number of spectral points and $\mathrm{MAX_D}$ denotes the peak value of the directly measured spectrum. The retrieved transmission spectra from SGI are shown in **Figures 4**a–**4**d with corresponding PSNR values of 22.84, 23.74, 24.47, and 24.64 dB for 16, 24, 32, and 64 measurements, respectively. These values confirm that the reconstruction accuracy improves with the number of probing patterns. Remarkably, the spectral features can already be resolved with only 16 patterns corresponding to a 25 % sampling ratio, which significantly reduces the measurement time.

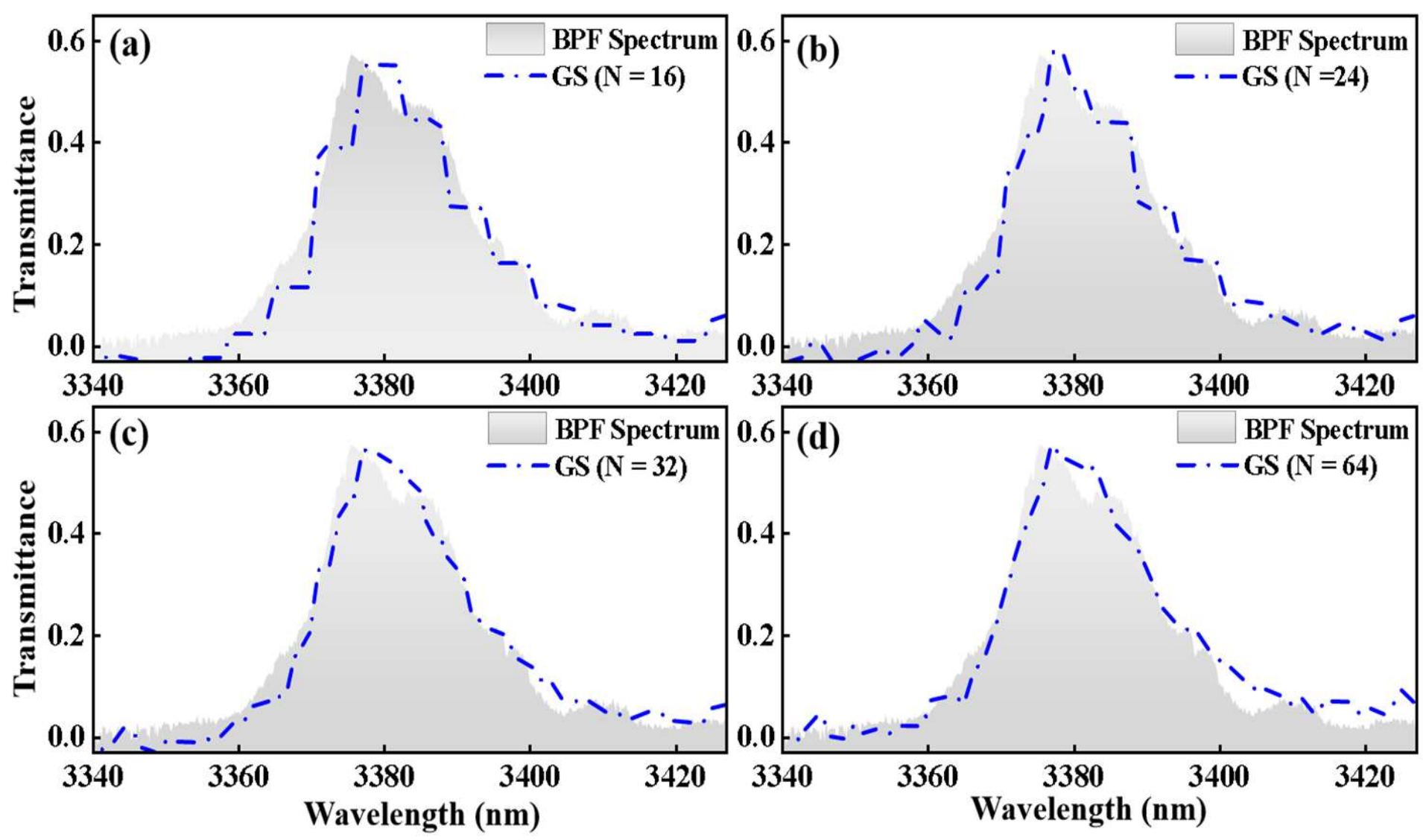

**Figure 4.** Comparison between retrieved ghost spectra and directly measured transmission of the bandpass filter (BPF, shaded area) with the optical spectrum analyzer for different numbers of probing spectral patterns using a 64-dimensional Hadamard matrix. (a–d) Retrieved ghost spectra (blue dashed lines) obtained with 16, 24, 32, and 64 probing spectral patterns, respectively.

After confirming the reconstruction accuracy and sampling efficiency of the proposed ghost spectroscopy approach, we next demonstrate its spectral versatility enabled by the frequency-down-conversion process. The operating wavelength of the computational ghost spectroscopy system can be flexibly tuned by adjusting either the spectral range of the 1.5 μm source or the wavelength of the 1 μm pump laser, allowing for the generation of broadband tunable MIR idler light centered at 3290, 3375, and 3432 nm as shown in **Figures 5**a–**5**c. The temperature of the CPLN crystal was precisely controlled with a heating stage to maintain phase-matching. The parameters for tunable MIR generation, including signal and pump wavelengths as well as the CPLN crystal temperature can be found in Note 2 of the Supplementary Material. We then applied the ghost spectroscopy scheme across the full 3.25–3.47 μm region to measure the transmission spectrum of a 0.05 mm-thick polystyrene film. Specifically, using a 64-dimensional Hadamard matrix to spectrally modulate the 1.5 μm light, the probing patterns were sequentially applied while tuning the idler across the three spectral windows centered at 3290, 3375, and 3432 nm. The retrieved ghost spectra (colored dotted lines) shown in Figure 5d clearly resolve the four distinct absorption peaks of polystyrene located at 3267, 3307, 3334, and 3420 nm in this region and closely match the spectrum directly measured with a commercial Fourier-transform infrared (FTIR) spectrometer (4 $cm^{-1}$ resolution, shaded region). These results demonstrate broadband and wavelength-tunable operation of computational ghost spectroscopy in the MIR.

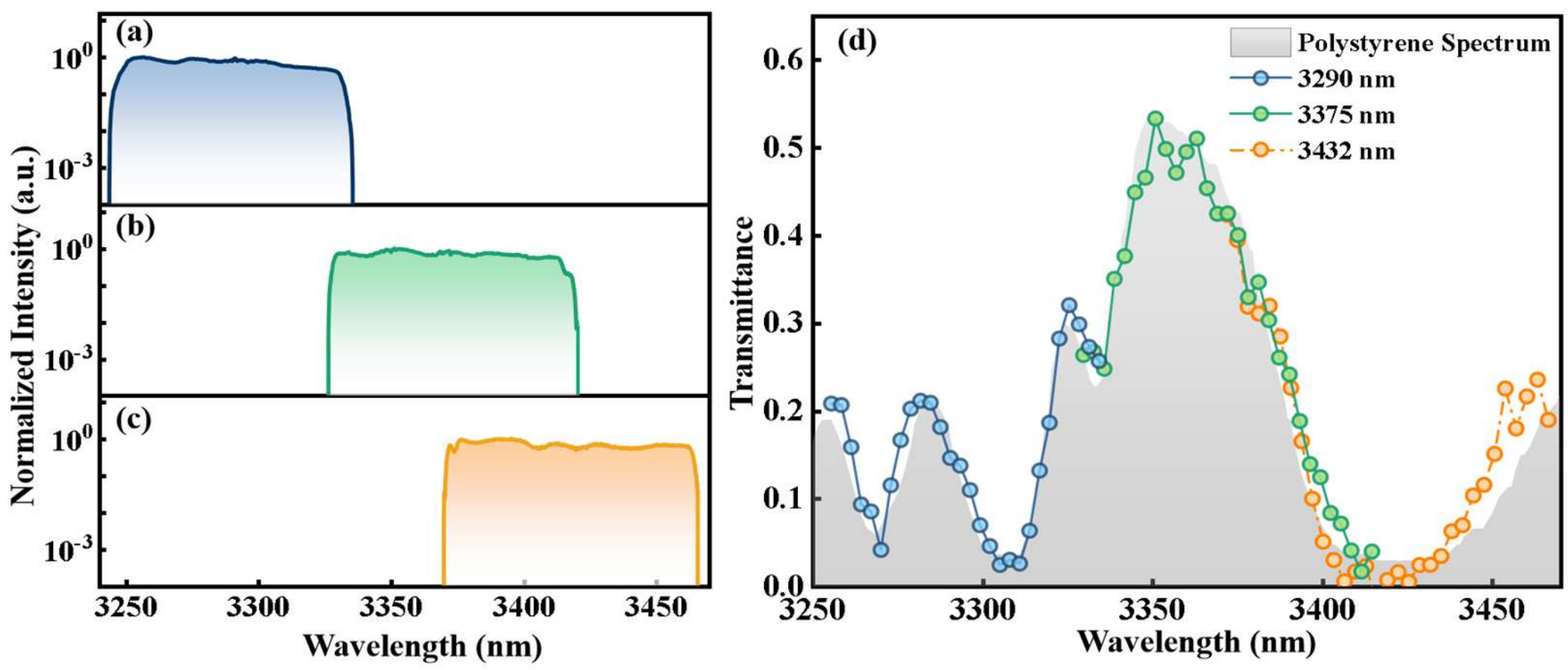

**Figure 5.** Tunable mid-infrared spectra generated through nonlinear frequency down-conversion and absorption measurement of a 0.05-mm polystyrene film using computational ghost spectroscopy. (a–c) Tunable mid-infrared idler spectra centered at 3290, 3375, and 3432 nm, respectively, generated by difference-frequency generation in a chirped-poling lithium niobate crystal. (d) Reconstructed transmission spectra (colored dotted lines) obtained by computational ghost spectroscopy compared with the reference spectrum measured by Fourier-transform infrared spectroscopy (shaded region) over the 3.25–3.47 μm range.

### 3.3 | Extension of computational ghost spectroscopy to the long-wavelength infrared

Building on the demonstrated wavelength tunability in the mid-infrared, we next explore the scalability of the approach toward longer wavelengths. The operational range of computational ghost spectroscopy using a CPLN crystal is limited to below 5 μm due to its transparency window. Extending the scheme into the molecular-fingerprint region beyond 5 μm can however be achieved by combining a wavelength-agile near-infrared fiber laser with nonlinear frequency down-conversion in a non-oxide crystal. As a proof of concept, we demonstrate computational ghost spectroscopy in the long-wavelength infrared beyond 7 μm using a zinc germanium phosphide (ZGP) crystal. ZGP provides a broad transmission window (0.74–12 μm) and a high nonlinear coefficient (~60 pm/V), enabling efficient frequency conversion and access to the long-wavelength infrared region where many molecular and biological absorption resonances are located[37,38].

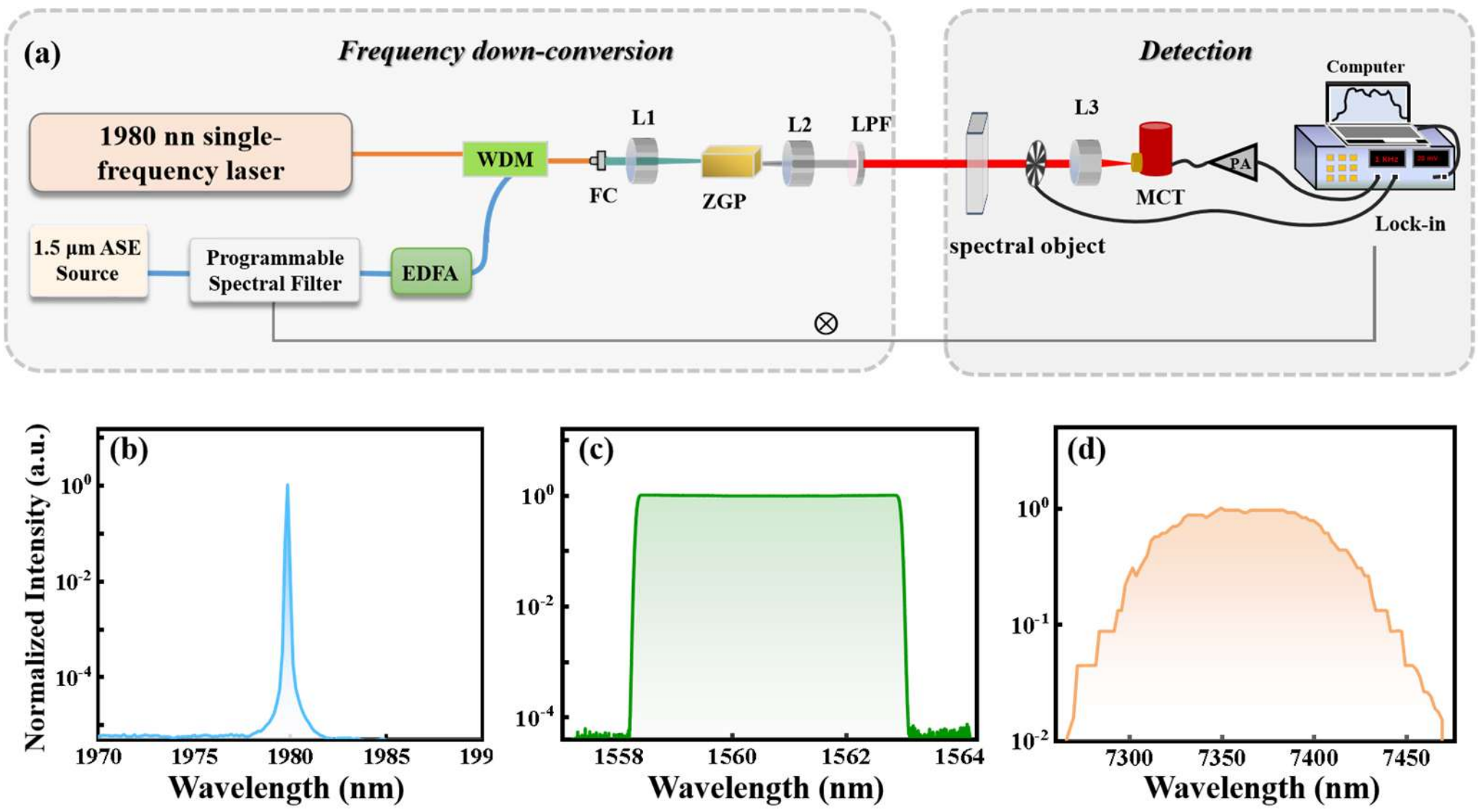

**Figure 6.** Schematic and spectral characteristics of the long-wavelength infrared computational ghost spectroscopy system. (a) Experimental setup based on frequency down-conversion in a zinc germanium phosphide (ZGP) crystal. (b) Spectrum of the single-frequency 1980 nm pump laser. (c) Spectrum of the filtered 1.5 μm signal used for difference-frequency generation, showing a 3 dB bandwidth of 4.8 nm. (d) Measured spectrum of the generated long-wavelength infrared idler, exhibiting a bandwidth of approximately 100 nm.

The schematic of the experimental setup is shown in **Figure 6**a. A ZGP crystal (6 × 6 × 4 mm) with anti-reflection coatings in the 1.5–2.2 μm and 6–10 μm bands was used. The crystal was cut at θ = 60.2° and φ = 0° for Type-I birefringent phase matching. By replacing the 1064 nm pump with a 1980 nm single-frequency laser (see spectrum in Figure 6b), an idler centered near 7 μm was generated through DFG in the ZGP crystal. The phase-matching bandwidth for the 1.5 μm signal was measured to be approximately 6 nm[37]. We then used a 32-dimensional Hadamard matrix to spectrally modulate the 1.5 μm light over a ~5 nm bandwidth. To characterize the idler generated in the ZGP crystal, it was coupled into a grating-scanning monochromator (Zolix Omni-λ500i) with ~20 nm resolution around 7 μm. As shown in Figures 6c and 6d, the 1.56 μm signal with 4.8 nm bandwidth was fully converted into a 7.35 μm idler with a −3 dB bandwidth of ~100 nm.

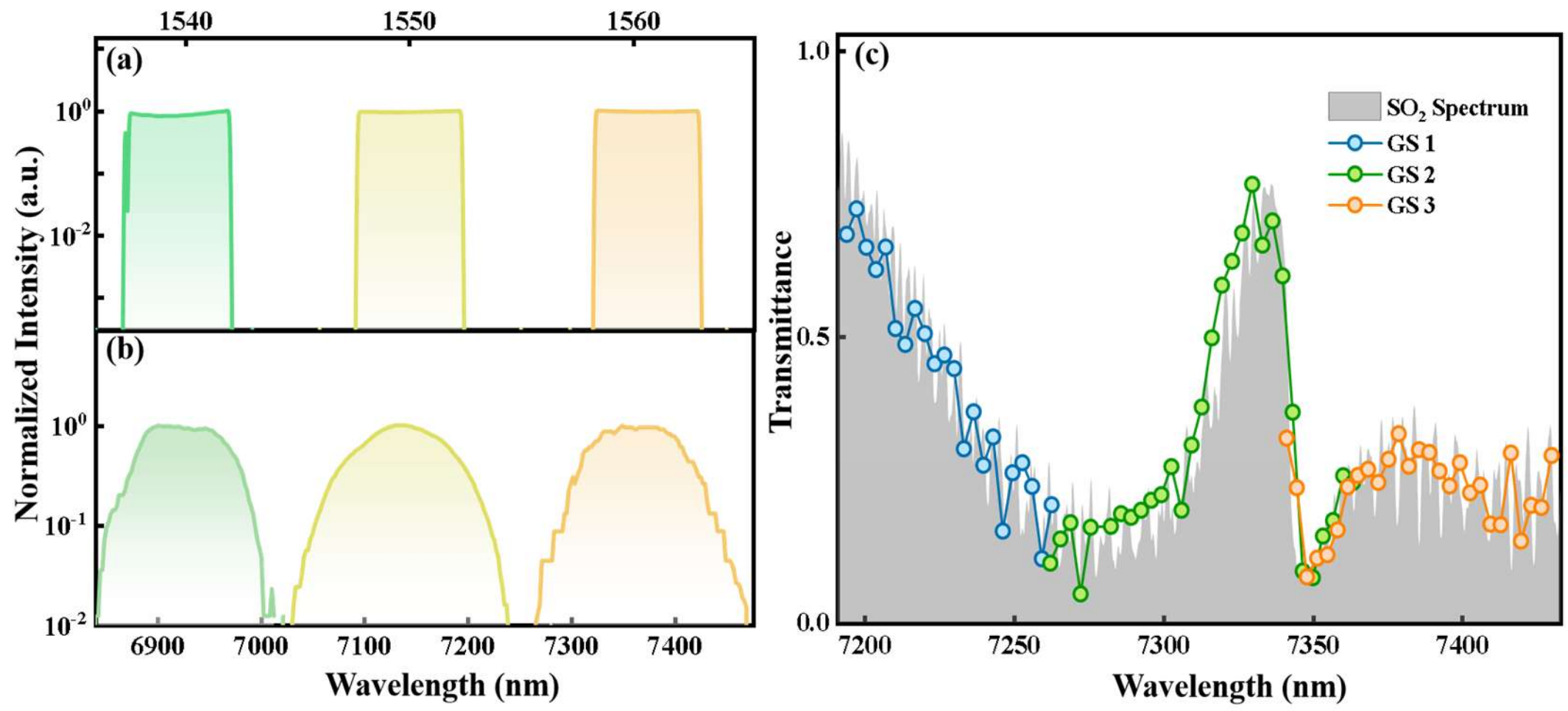

**Figure 7.** Wavelength tunability of the long-wavelength infrared computational ghost spectroscopy system and sulfur dioxide spectroscopy results. (a) Spectrally tuned 1.5-μm signal ranging from 1540 to 1560 nm. (b) Corresponding idler spectra generated via difference-frequency generation in the zinc germanium phosphide crystal, covering 6.9–7.45 μm. (c) Measured $SO_2$ transmission spectra: colored dotted lines represent the reconstructed ghost spectra, in close agreement with the reference data from the HITRAN database (shaded area).

We next characterize the wavelength tunability by fixing the 1980 nm pump and tuning the central wavelength of the 1.5 μm signal from 1540 nm to 1560 nm [**Figure** 7a] while adjusting the crystal angle to maintain phase matching. As shown in Figure 7b, the generated idler spectra cover the 6.9–7.45 μm range, consistent with the predicted phase-matching behavior of the ZGP crystal. The wavelength tuning confirms the capability of the system to access different portions of the long-wavelength infrared spectral region. The spectrally modulated 7 μm idler was then passed through an AR-coated long-pass filter to block residual pump and signal light and then transmitted through a target spectral object in the form of sulfur dioxide ($SO_2$) gas filled in a 10 cm long gas cell. The transmitted intensity was detected using a liquid-nitrogen-cooled HgCdTe detector (Judson DMCT16-De01) and recorded with a lock-in amplifier. The ghost spectrum was reconstructed in the 7.2–7.45 μm range [Figure 7c] from the second-order correlation between the measured intensities and the 32-dimensional

Hadamard matrix. One can see how the reconstructed spectra (colored dotted lines) closely match the reference $SO_2$ absorption spectrum (shaded area) from the HITRAN database (at 296 K and 1013.15 mbar). Additionally, we also achieved ghost spectroscopy over the 6.9 – 7.2 μm range by tuning the pump/signal wavelength and the crystal phase-matching angle (see Note 3 in supplementary Material). These results extend computational ghost spectroscopy into the long-wavelength infrared using a single-pixel detector.

### 3.4 | Comparison with conventional wavelength-scanning spectroscopy

An alternative method for realizing MIR spectroscopy involves wavelength-scanning, which applies narrowband tunable spectral filtering to near-infrared (NIR) light and detects the transmitted power of the MIR light after it passes through the spectral object. Compared with this conventional wavelength-scanning approach, computational ghost spectroscopy offers two main advantages. First, wavelength-scanning spectroscopy requires a 100% sampling ratio. In contrast, by employing preprogrammed spectral patterns, as we have already demonstrated in Fig.4, the sampling ratio of computational ghost spectroscopy can be significantly reduced. In fact, the spectral object can be resolved at only a 25% sampling ratio, which accordingly shortens the measurement time relative to the wavelength-scanning method. Second, computational ghost spectroscopy exhibits stronger resilience under weak-light conditions. This is because the transmitted power after the spectral object simultaneously contains contributions from all spectral components, and the correlation-based reconstruction inherently suppresses uncorrelated noise. Conversely, in wavelength-scanning spectroscopy, the transmitted powers at certain specific wavelengths may fall below the detector's response threshold under weak-light conditions, and detector noise along with environmental fluctuations can severely degrade the quality of spectral recovery.

To experimentally verify the superior performance of our computational ghost spectroscopy under weak-light conditions, we conducted a comparative experiment with conventional wavelength-scanning spectroscopy under identical spectral coverage, optical power, and detection conditions. In the SGI scheme, the programmable spectral filter encoded the broadband 1.5-μm source over the range of 1556.472–1558.392 nm using 16-bit complementary patterns, with a spectral resolution of 15 GHz. The encoded patterns were transferred to the MIR region via difference-frequency generation (DFG) in the CPLN crystal, and the light transmitted through a 10-cm methane gas cell (1% concentration) was recorded by a single-pixel MIR detector. For the wavelength-scanning method, the broadband 1.5-μm source was spectrally filtered by the same waveshaper, with the tunable central wavelength sweeping from 1556.472 to 1558.392 nm and the same spectral resolution of 15 GHz. The transmitted intensity of the tunable MIR light after DFG was then recorded point by point using the same detector.

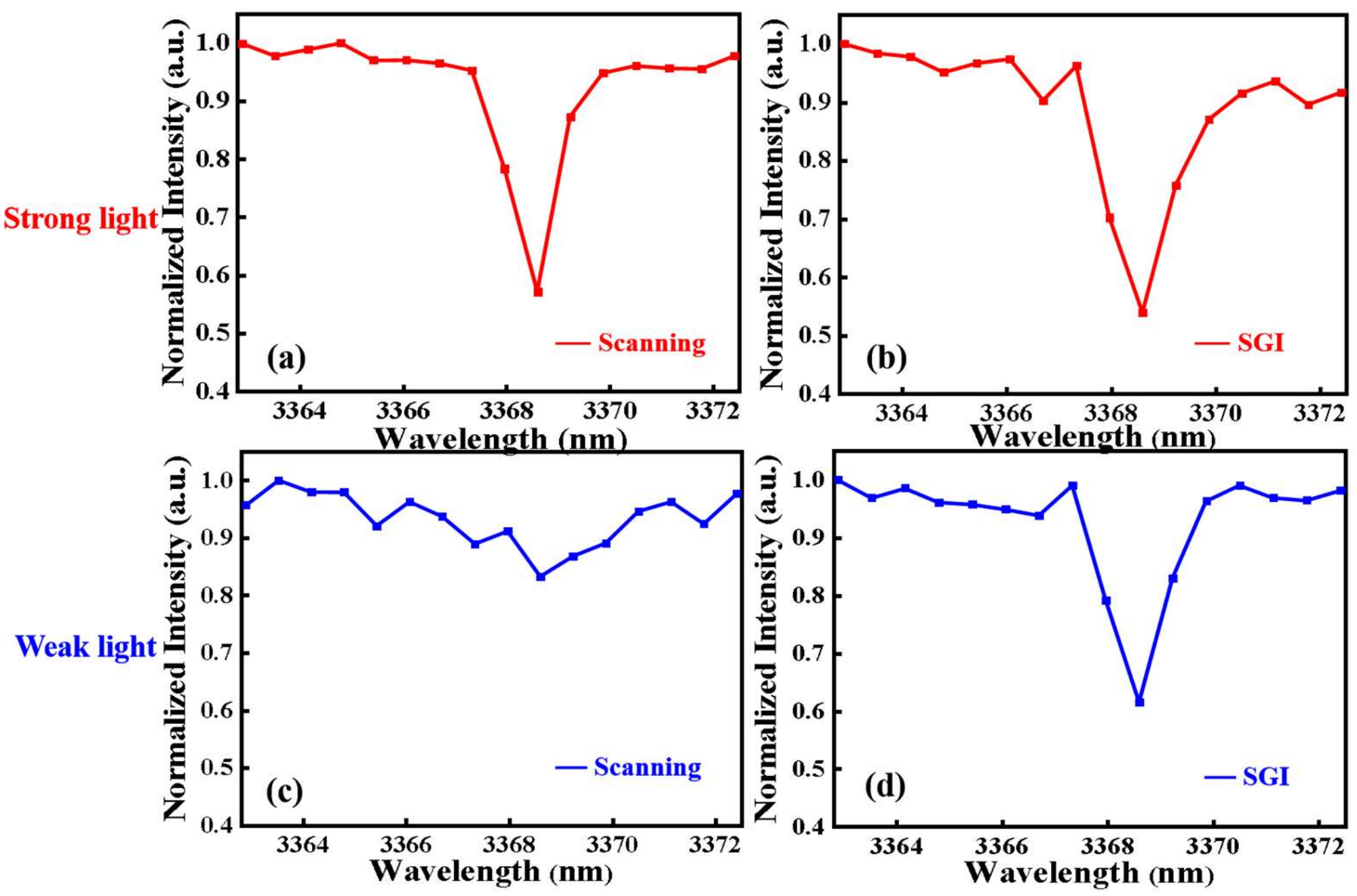

**Figure 8.** Comparison of methane absorption spectra obtained by conventional wavelength-scanning spectroscopy and computational ghost spectroscopy (GS) under strong- and weak-light conditions. (a) Strong-light wavelength-scanning spectroscopy. (b) Strong-light SGI. (c) Weak-light wavelength-scanning spectroscopy. (d) Weak-light SGI. The ghost spectroscopy scheme demonstrates clear advantage in the low-flux regime, successfully recovering the 3368.6 nm absorption feature where scanning spectroscopy fails.

The comparison was performed under two MIR power regimes: a relatively strong-light condition with 80 μW MIR power and a weak-light condition with 2 μW MIR power. As shown in **Figure** 8a and 8b, under strong illumination, both methods successfully resolved the methane absorption feature around 3368.6 nm with comparable quality. However, under weak-light conditions (**Figure** 8c and 8d), the scanning method failed to recover the absorption peak, whereas the computational ghost spectroscopy scheme was still able to reconstruct the methane absorption feature. These results clearly demonstrate the distinct advantage of our computational ghost spectroscopy for weak-light MIR applications, such as spectroscopic measurements of highly absorbing samples or remote sensing in strongly scattering environments.

## 4 | Discussion

We have demonstrated computational ghost spectroscopy in the mid-infrared region through nonlinear frequency down-conversion. To address the challenge of directly programming spectral patterns in this spectral range, predesigned Hadamard patterns were applied to a broadband 1.5 μm source using a programmable spectral filter and subsequently transferred to the mid-infrared via difference-frequency generation in nonlinear crystals. The concept was validated experimentally by achieving broadband operation over 3.25–3.47 μm with a spectral resolution of 0.62 $cm^{-1}$ (corresponding to 0.71 nm at ~3380 nm) using a chirped-poling lithium niobate crystal. By combining wavelength-agile near-infrared fiber lasers with non-oxide nonlinear crystals, we further extended the approach to the long-wavelength infrared region, demonstrating gas spectroscopy between 7.2 and 7.45 μm using a zinc germanium phosphide crystal. This method removes the need for high-sensitivity mid-infrared detector arrays, inheriting the intrinsic advantages of ghost imaging and offering strong resilience under weak-light and turbulent conditions, which is essential for open-path and harsh-environment applications.

Several routes exist to enhance the performance of the technique further. The accessible spectral range could be extended by employing widely tunable fiber lasers such as ytterbium-doped sources (1.03–1.09 μm) or thulium-doped sources (1.8–2.0 μm), enabling operation across the 3–3.8 μm and 6.5–11.5 μm regions, respectively. Broadband long-wavelength infrared ghost spectroscopy spanning 9–11 μm could be achieved using $BaGa_4Se_7$ crystals[39], which provide much broader phase-matching bandwidth for 1.5 μm pumping (see Note 4 in the Supplementary Material). Meanwhile, employing $LiGaS_2$ (LGS) crystal under 1 μm pumping, a broadband ghost spectroscopy spanning approximately 7 – 9 μm can be realized (see Note 5 in supplementary Material). The current acquisition rate, limited by the switching time of the programmable spectral filter (~1 s), could be increased by several orders of magnitude through spectral–temporal modulation based on dispersive Fourier transform techniques[40, 41], enabling acquisition rates in the tens of kHz range.

Further improvements in spectral resolution are also possible using e.g. programmable electro-optic frequency combs at 1.5 μm[42,43] as the modulated source in the down-conversion process, yielding potential resolution at the MHz level. In addition, implementing up-conversion detection[44, 45] would permit detection in the visible or near-infrared using silicon or indium-gallium-arsenide single-photon detectors, thereby enhancing sensitivity and reducing noise.

Our work establishes a versatile and scalable framework for infrared spectroscopy that combines high resolution, speed, sensitivity, wavelength tunability, and environmental robustness. The proposed approach opens new opportunities for next-generation mid- and long-wavelength infrared sensing, particularly for remote detection and molecular-fingerprint spectroscopy in complex environments.

## Supplementary Information

The online version contains supplementary material available at Springer Open.

**Declarations**

**Availability of data and materials**

The data that support the plots and maps within this paper and other findings are available from the corresponding authors upon reasonable request.

**Competing interest**

The authors declare no competing interests.

**Funding**

This work was supported by the National Natural Science Foundation of China (Grant No. 62375189, W2411056, U22A2090), the Postdoctoral Fellowship Program of CPSF (Grant No. GZB20240472, 2025M770814) and the Research Council of Finland (Grant No. 368650)

**Authors' contributions**

H.W conceived the idea of this work. H.W., G.G., H.L. and W. Z. designed the experiment. L.H, H.W., M.S., Z.B., B.H., and W.D. built the experimental setups and carried out all the experiments. H.W., L.H., G.G., S.H. and H. L. prepared the manuscript. All the authors discussed the results and contributed the paper.

**Acknowledgments**

The authors thank Dr. Yani Xie from the Analysis and Testing Center of Sichuan University for helping us perform IR studies.

**Authors' information**

Corresponding authors:

Han Wu, Email: hanwu@scu.edu.cn

Houkun Liang, Email: hkliang@scu.edu.cn

Weili Zhang, E-mail: wl_zhang@uestc.edu.cn

Goëry Genty, E-mail: goery.genty@tuni.fi